\newcommand {\bea}{\begin{eqnarray}}
\newcommand {\eea}{\end{eqnarray}}
\newcommand {\be}{\begin{equation}}
\newcommand {\ee}{\end{equation}}

\documentstyle[12pt]{article}
\begin{document}
{\hbox to\hsize{November 1998 \hfill IASSNS-HEP 98/100}}\par
\begin{center}
{\LARGE \bf Continuity of Quark and \\[0.1in] 
Hadron Matter} \\[0.3in]
{\bf Thomas Sch\"afer}\footnote{Research supported in part by NSF
PHY-9513835.\\  e-mail: schaefer@sns.ias.edu}\\[.05in]
and \\ [.1in] 
{\bf Frank Wilczek}\footnote{Research supported in part by DOE grant
DE-FG02-90ER40542.\\  e-mail: wilczek@sns.ias.edu } \\[.05in]
{\it School of Natural Sciences\\
Institute for Advanced Study\\
Princeton, NJ 08540}\\[.15in]
\end{center}
\bigskip
\bigskip
\bigskip
\begin{abstract}

We review, clarify, and extend the notion of color-flavor
locking.  We present evidence that for three degenerate flavors the
qualitative features of the color-flavor locked state, reliably
predicted for high density, match the expected features of hadronic
matter at low density.  This provides, in particular, a controlled,
weak-coupling realization of confinement and chiral symmetry breaking
in this (slight) idealization of QCD.

\end{abstract}

\newpage

In a recent study \cite{ARW_98b} of QCD with three degenerate flavors 
at high density, a new form of ordering was predicted, wherein the color 
and flavor degrees of freedom become rigidly correlated in the 
groundstate: color-flavor locking.  This prediction is based on a weak
coupling analysis using a four-fermion interaction with quantum
numbers abstracted from one gluon exchange.  One expects that such a
weak coupling analysis is appropriate at high density, for the
following reason \cite{BL_84,ARW_98a,RSSV_98}. Tentatively assuming 
that the quarks start out in a state close their free quark state, 
{\it i.e}. with large Fermi surfaces, one finds that the relevant 
interactions, which are scatterings the states near the Fermi surface, 
for the most part involve large momentum transfers. Thus, by asymptotic 
freedom, the effective coupling governing them is small, and the starting
assumption is confirmed.

Of course, as one learns from the theory of superconductivity 
\cite{S_64}, even weak couplings near the Fermi surface can have 
dramatic qualitative effects, fundamentally because there are many 
low-energy states, and therefore one is inevitably doing highly 
degenerate perturbation theory. Indeed, the authors of \cite{ARW_98b} 
already pointed out that their color-flavor locked state, which is 
constructed by adapting the methods of superconductivity theory to 
the problem of high-density quark matter, displays a gap in all 
channels except for those associated with derivatively coupled spin 
zero excitations, {\it i.e}. Nambu-Goldstone modes. This is confinement.
For massless quarks, they also demonstrated spontaneous chiral symmetry 
breaking.

In very recent work we \cite{SW_98}, and others \cite{EHS_98}, have 
reinforced this circle of ideas by analyzing renormalization of the 
effective interactions as one integrates out modes far from the Fermi 
surface. A fully rigorous treatment will have to deal with the extremely
near-forward scatterings, which are singular due to the absence of
magnetic mass for the gluons, at least in straightforward perturbation
theory.  This problem, which is presumably technical, is any case
ameliorated self-consistently for states of the color-flavor locking
type, wherein all the gluons acquire mass through the Anderson-Higgs
mechanism.

In the earlier work \cite{ARW_98b}, several striking analogies between 
the calculated properties of the color-flavor locked state and the
expected properties of hadronic matter at low or zero density, based 
on standard lore and observed phenomenology, were noted.  In addition
to confinement and chiral symmetry breaking, the authors observed that
the dressed elementary excitations in the color-flavor locked state
have the spin quantum numbers of low-lying hadron states and for the
most part carry the expected flavor quantum numbers, including
integral electric charge (in units of the electron charge).  Thus, as
we shall spell out immediately below, the gluons match the octet of
vector mesons, the quark octet matches the baryon octet, and an octet
of collective modes associated with chiral symmetry breaking matches
the pseudoscalar octet.  However there are also a few apparent
discrepancies: there is an extra massless singlet scalar, associated
with the spontaneous breaking of baryon number (superfluidity); there
are eight rather than nine vector mesons (no singlet); and there are
nine rather than eight baryons (extra singlet).  We will argue that
these ``discrepancies'' are superficial -- or rather that they are
features, not bugs.

Let us first briefly recall the fundamental concepts of color-flavor
locking.  The case of three massless flavors is the richest due to its 
chiral symmetry (and adding a common mass does not change anything 
essential) so we shall concentrate on it. The primary condensate, 
which one calculates using the methods of superconductivity theory 
near the Fermi surface, involves diquarks. It takes the 
form \cite{ARW_98b}
\be
\label{cfl}
\langle q_{La}^{i\alpha}
q_{Lb}^{j\beta} \epsilon_{ij} \rangle = -\langle q_{R\dot k a}^{\alpha}
q_{R\dot l b}^{\beta} \epsilon^{\dot k \dot l} \rangle = \kappa_1
\delta^\alpha_a \delta^\beta_b + \kappa_2 \delta^\alpha_b
\delta^\beta_a 
\ee
Here $L$, $R$ label the helicity, $i,j,k,l$ are two-component spinor
indices, $a,b$ are flavor indices, and $\alpha, \beta$ are color
indices.  A common space-time argument is suppressed.  $\kappa_1,
\kappa_2$ are parameters (depending on chemical potential, coupling,
$\ldots$) whose non-zero values emerge from a dynamical calculation.

This equation must be interpreted carefully.  The value of any local
quantity which is not gauge invariant, taken literally, is meaningless, 
since local gauge invariance parameterizes the redundant variables in 
the theory, and cannot be broken \cite{R_75}. But as we know from
the usual treatment of the electroweak sector in the Standard Model,
it can be very convenient to use such quantities.  The point is that
we are allowed to fix a gauge during intermediate stages in the
calculation of meaningful, gauge invariant quantities -- indeed, in
the context of weak coupling perturbation theory, we must do so.  For
our present purposes however it is important to extract
non-perturbative results, especially symmetry breaking order
parameters, that we can match to our expectations for the hadronic
side.  To do this, we can take suitable products of the members of
(\ref{cfl}) and their complex conjugates, and contract the color
indices.  In this way we can produce the square of the standard chiral
symmetry breaking order parameter of type
$\langle \bar {q_L} q_R \rangle$ and a baryon number violating order
parameter of type $\langle (qqq)^2 \rangle$, both scalars and singlets
under color and flavor.  At this level only the square of the usual
chiral order parameter appears, fundamentally because our condensates
preserve left-handed quark number modulo two.  This conservation law
is violated by the six-quark vertex associated with instantons, and by
convolving that vertex with our four-quark condensate we can obtain
the usual two-quark chiral symmetry breaking order parameter
\cite{RSSV_98b}.

By demanding invariance of the diquark condensate directly, we infer
the symmetry breaking pattern $SU(3)^c \times SU(3)^L \times SU(3)^R
\times U(1) \rightarrow SU(3)^\Delta$.  Here among the initial
microscopic symmetries $SU(3)^c$ is local color symmetry, while the
remaining factors are chiral family and baryon number symmetries.  The
final residual unbroken symmetry is a global diagonal symmetry.
Indeed, the Kronecker deltas in the final term of (\ref{cfl})  are
invariant only under simultaneous color and flavor rotations, so the
color and flavor are ``locked''.  This locking occurs separately for the
left and right handed quarks, but since color symmetry itself is
vectorial, the effect is also to lock left and right handed flavor
rotations, breaking chiral symmetry.  The global baryon number
symmetry is, of course, manifestly broken, but quark number is
conserved modulo two.  Projecting onto the gauge invariant, color
singlet, sector this implies that baryon number is violated only
modulo two.  The same conclusions would emerge from analysis of the
gauge invariant symmetry generators only, upon consideration of the
gauge invariant order parameters we constructed above.

Ordinary electromagnetic gauge invariance, like color symmetry, is
violated by (\ref{cfl}), but a linear combination of hypercharge
(diagonal matrix -2/3, 1/3, 1/3) and electromagnetic charge (diagonal
matrix 2/3, -1/3, -1/3) annihilates the combinations correlated by
color-flavor locking, and generates a true symmetry.  The physical
result is that there is a massless gauge degree of freedom,
representing the photon as modified by its interaction with the
condensate.  As seen by this modified photon, all the elementary
excitations have appropriate charges to match the corresponding
hadronic degrees of freedom.  In particular, their charges are all
integral multiples of the electron charge \cite{ARW_98b}.  This is, 
of course, another classic aspect of confinement.   

It was essential, in this construction, that the charges of the quarks
add up to zero.   If that were not so, we would not have been able to
find a color generator capable of compensating the violation of naive
electromagnetic gauge invariance.  Yet it seems somewhat accidental
that these charges do add up to zero, and one would be quite worried
if any qualitative aspect of confinement depended on this
accident. This worry touches the form rather than the substance of our
argument. If the quark charges did not up to zero, it would not be
valid to ignore Coulomb repulsion.  One would have to add  a
compensating charge background as a mathematical device, or
contemplate inhomogeneous states.  Insofar as we want to use external
gauge fields as a probe of pure QCD, we must restrict ourselves to
those which preserve the overall neutrality of the QCD groundstate.
Fortunately, in our slightly idealized version of QCD no awkwardness
arises for the physically important gauge field, {\it i.e}. the
physical photon.

The elementary excitations are of three types.  The color gluons
become massive vector mesons through the Anderson-Higgs mechanism.
Due to color-flavor locking, they acquire flavor quantum numbers,
which makes them an octet under the residual $SU(3)^\Delta$.  The
quark fields give single-particle spin 1/2 excitations whose stability
is guaranteed by the residual $Z_2$ quark (or baryon) number symmetry.
These excitations are massive, due to the color-flavor superconducting
gap.  They form an octet with the quantum numbers of the nucleon
octet, plus a singlet.  It might seem peculiar on first hearing that a
single quark can behave as a baryon, but remember that there is a
condensate of diquarks pervading this phase.  In addition there are
collective Nambu-Goldstone modes, associated with the spontaneously
broken global symmetries.  These are a massless pseudoscalar octet
associated with chiral symmetry breaking, and a scalar singlet
associated with baryon number violation.  A common quark mass lifts
the pseudoscalar octet, but not the singlet, because it spoils
microscopic chiral symmetry but not microscopic baryon number.

Clearly, there are striking resemblances between the elementary
excitations of color-flavor locked quark matter and the low-energy
hadron spectrum.  One is tempted to ask whether they might be
identified.  More precisely, one might ask whether strongly coupled
hadronic matter at low density goes over into the calculable,
weak-coupling form of quark matter just described without a phase
transition.  If so, then the confinement and chiral symmetry breaking
calculated for the weak coupling calculation not only resemble these
central properties of low-density QCD, but are rigorously
indistinguishable from them.  This sort of possibility, that Higgs and
confined phases are rigorously indistinguishable, has long been known
to occur in simple abstract models \cite{FS_79}.

As mentioned above, however, at first sight there appear to be several
difficulties with this identification.  We now debunk them in turn.

The most profound of the apparent difficulties is the existence of an
extra scalar Nambu-Goldstone mode, and the related phenomenon that
baryon number is spontaneously violated (indicating, as in liquid
He$^4$, superfluidity).  The answer to this comes through proper
recognition of an important though somewhat exotic phenomenon for
three degenerate flavors on the hadron side.  Several years ago
R. Jaffe discovered \cite{J_77}, in the context of the MIT bag model, 
that a particular dibaryon state, the H, a spin 0 SU(3) singlet with 
quark content (udsuds), is surprisingly light.  This arose, in his
calculations, because of a particularly favorable contribution from
color magnetism.  Roughly speaking, in the H configuration the color
fields associated with the quark sources are minimized, together with
the energy they would otherwise store, by arranging both the colors
and spins to cancel pairwise to the greatest extent possible.  It has
been debated, for QCD with realistic quark masses, whether H might be
only slightly above the nn or n$\Lambda$ thresholds. Though at
this level the outcome for realistic QCD is unclear, both
theoretically \cite{D_89} and experimentally \cite{Conf_98}, it 
has come to seem quite likely that in QCD with three degenerate 
quarks the H will be the particle with smallest energy per unit 
baryon number. Thus at any finite baryon number density, however 
small, at zero temperature one should expect, in this context, 
to find a Bose condensate of H dibaryons.  This condensate gives us
precisely -- {\it i.e}. with the appropriate quantum numbers -- the
superfluid we were led to expect from our superficially very different
considerations on the quark matter side.  

If the H is above dibaryon threshold, one will have a narrow range of
chemical potentials where baryon number is built up by single
baryons.  Based on the same calculations \cite{J_77,D_89}, it 
is extremely plausible that in this case there will be attraction 
in the H channel at the Fermi
surface, and hence superfluidity of the required type, now 
through a BCS mechanism.

This superfluidity, whatever its source, 
supplies us with the key to the riddle of the missing vector
meson.  For once there is a massless singlet scalar, the putative
singlet vector becomes radically unstable, and should not appear in
the effective theory.  It might be objected here that the octet of
vector mesons is also unstable -- for massless quarks -- against decay
into massless scalar and pseudoscalar mesons.  A quick answer is that
this is not really an objection at all, because there is no harm in
having redundant states (whose instability will appear immediately
upon more accurate calculation).  There is a much prettier and more
satisfying answer, however.  If we turn on non-zero masses for the
quarks the pseudoscalar octet (but not the singlet) will become
massive.  Eventually the decay of the vector octet (but not the
singlet) will be blocked, and then we will be grateful for the
prescience of the theory in providing the appropriate degrees of
freedom.

Finally, there is the question of the ``extra'' singlet baryon.  This is
the most straightforward.  In the original calculations \cite{ARW_98b}, 
it was found that the singlet gap is much larger than the octet gap.  
Thus the singlet baryon is predicted to be considerably heavier than the
octet. This is not problematic: a particle of this sort is expected
in the quark model, it could well exist in reality, and in any case it
is radically unstable against decay into octet baryon and octet
pseudoscalar, at least for massless or light quarks.

So all the objections have been answered.  Continuity of quark and
hadron matter, far from being paradoxical, now appears as the default
option.

Clearly, superfluidity of quark/hadron matter has been essential for
the argument. There is considerable evidence for pairing in nuclei
\cite{RS_80}. Its full realization is limited by the finite size of 
nuclei, which in turn arises from the non-negligible strange quark 
mass and the Coulomb energy that arises in the most favorable (for QCD), 
symmetric arrangement of neutrons and protons. These limitations might be
relieved to some extent in heavy ion collision accompanied by creation
of many strange-antistrange pairs, followed by charge segregation.  An
important signature for this, emphasized by the considerations above,
is broadening of vector mesons, especially the singlet.  This effect
might be observable in the dimuon spectrum.

Our considerations here are clearly relevant to any attempt to model
the deep interior of neutron stars, or conditions during supernova and
hypernova explosions.  To do justice to these questions, it will be
very important to include the effects of unequal quark masses and of
electromagnetism.  That is an important task for the future.

In the remainder of this paper we shall consider a related but simpler
problem, that of extending the analysis to larger numbers of
degenerate quarks.  An important foundational result, which emerges
clearly from this analysis, is that the color-flavor locked state for
three flavors, which was first guessed to be favorable because of its
large residual symmetry and by analogy to the B phase of superfluid
He$^3$, is in fact the global minimum for three flavors.  It also
reappears as a building block for larger numbers of flavors.

The renormalization group analysis in \cite{SW_98,EHS_98} allows one 
to classify possible instabilities, and to assess their relative 
importance, for small but otherwise arbitrary couplings near the 
Fermi surface. It was found that the dominant instability
corresponds to scalar diquark condensation. The analysis
does not fix the color and flavor channel of this instability
uniquely, independent of initial conditions for the couplings, since
there are two equally enhanced marginal interactions.  
One gluon exchange, which dominates for weak coupling,
is attractive in the color anti-symmetric $\bar 3$ channel, and 
favors one of these interactions.  During the
evolution this interaction will grow, while the repulsive
interaction in the color symmetric 6 channel is suppressed.
Thus the instability is driven by a leading interaction of the
form
\bea
\label{laa}
 {\cal L} &=& K
  \left(\delta_{\alpha\gamma}\delta_{\beta\delta}-\delta_{\alpha\delta}
\delta_{\beta\gamma}\right)
  \left(\delta^{ac}\delta^{bd}-\delta^{ad}\delta^{bc}\right)
  \nonumber \\
& &\hspace{2cm} \left\{ \left(\psi^\alpha_a C\gamma_5 \psi^\beta_b \right)
     \left(\bar\psi^\gamma_c C\gamma_5 \bar\psi^\delta_d \right)
     - \left( C\gamma_5\leftrightarrow C\right )\right\},
\eea
where as before $\alpha, \beta, \ldots$ 
are color indices and $a, b, \ldots$ are flavor
indices. The Dirac structure of the interaction becomes more
transparent when written in a chiral basis. We have
\bea
\label{LLLL}
 \epsilon_{ij}\epsilon_{kl}
 \psi_L^i\psi_L^j \bar\psi_L^k\bar\psi_L^l
 + (L\leftrightarrow R).
\eea

The renormalization group analysis only provides the form
of the dominant interaction, not the structure of the order
parameter. In particular, it does not tell us whether 
color-flavor locking is the preferred state in three
flavor QCD. In order to answer this question, we have to
perform a variational analysis. Since the interaction is
attractive in s-wave states, it seems clear that the 
dominant order parameter is an s-wave, too. We then 
only have to study the color-flavor structure of the 
primary condensate. For this purpose, we calculate the 
effective potential for the order parameter
\bea
\label{order}
 \langle \psi_{L\,a}^{i \alpha}\psi_{L\,b}^{j \beta}\rangle
 = \epsilon^{ij}\Delta^{\alpha\beta}_{ab}.
\eea
$\Delta^{\alpha\beta}_{ab}$ is a $N_f\times N_f$ matrix in flavor
space and a $N_c\times N_c$ matrix in color space. Overall
symmetry requires that $\Delta^{\alpha\beta}_{ab}$ is symmetric
under the combined exchange $(a \alpha)\leftrightarrow (b \beta)$.
Also, since the interaction only involves color and
flavor anti-symmetric terms, the effective potential does
not depend on color and flavor symmetric components of
$\Delta^{\alpha\beta}_{ab}$. This means that the effective potential
has at least $N_c(N_c+1)N_f(N_f+1)/4$ flat directions.
These trivial flat directions will be lifted by subleading
interactions not included in our analysis. We will comment
on the importance of subleading terms below.

   We calculate the effective potential in the mean field
approximation. This approximation corresponds to resumming
all ``cactus'' diagrams. These diagrams are expected to be dominant
both in the limit of large chemical potential and in the large
$N_c, N_f$ limit. In the mean field approximation, the
quadratic part of the action becomes
\bea
 {\cal M}^{\alpha\beta}_{ab}  \psi_{L\,\alpha}^a\psi_{L\,\beta}^b
 =K\left( \Delta^{\alpha\beta}_{ab} - \Delta^{\beta\alpha}_{ab}
   - \Delta^{\alpha\beta}_{ba} + \Delta^{\beta\alpha}_{ba} \right)
   \psi_{L\,\alpha}^a\psi_{L\,\beta}^b.
\eea
Integrating over the fermion fields we obtain the familiar
${\rm tr}\log$ term in the effective potential. In order to
evaluate the logarithm, we have to diagonalize the mass matrix
${\cal M}$. Let us denote the corresponding eigenvalues by
$\delta_\rho\,(\rho=1,\ldots,N_cN_f)$. These are the physical
gaps for the $N_fN_c$ fermion species. Adding the mean field
part of the effective potential, we finally obtain
\bea
\label{veff}
 V_{eff}(\Delta) = -\sum_{\rho} \epsilon(\delta_\rho)
  + {\cal M}_{\alpha\beta}^{ab}\Delta^{\alpha\beta}_{ab}.
\eea
Here, $\epsilon(\delta)$ is the kinetic term in the effective
action for one fermion species,
\bea
 \epsilon(\delta) = \int\frac{d^3p}{(2\pi)^3}\left(
 \sqrt{(p-\mu)^2+\delta^2}+\sqrt{(p+\mu)^2+\delta^2}\right).
\eea
This integral has an ultra-violet divergence. This divergence
can be removed by expressing $\delta$ in terms of the renormalized
interaction \cite{Wei_94}. In this work we are not really interested in
the exact numerical value of the gap, but only in the symmetries
of the order parameter. For simplicity, we therefore regularize the
integral by introducing a sharp three-momentum cutoff $\Lambda$.

 The effective potential (\ref{veff}) depends on $N_c(N_c-1)
N_f(N_f-1)/4$ parameters. We minimize this function numerically.
In order to make sure that the minimization routine does not
become trapped in a local minimum we start the minimization from
several different initial conditions. For the numerical analysis
we also have to fix the value of the chemical potential $\mu$,
the coupling constant $K$, and the cutoff $\Lambda$. We have
checked that the symmetry breaking pattern does not depend
on the values of these parameters. We have used $\mu=0.5$ GeV,
$\Lambda=0.6$ GeV and $K=3.33/\Lambda^2$, similar to what was
considered in \cite{ARW_98a,RSSV_98}.

  After we determine the matrix $\Delta^{\alpha\beta}_{ab}$ that minimizes
the effective potential we study the corresponding symmetry breaking.
Initially, there are $N_f^2-1$ global flavor symmetries for both
left and right handed fermions, as well as $N_c^2-1$ local gauge
symmetries. Superfluidity reduces the am-mount of symmetry. In order
to find the unbroken generators we study the second variation of the
order parameter $\delta^2\Delta/(\delta\theta_i\delta\theta_j)$, where
$\theta_i\;(i=1,\ldots,N_f^2+N_c^2-2)$ parameterizes the flavor and
color transformations. Zero eigenvalues of this matrix correspond
to unbroken color-flavor symmetries. The corresponding eigenvectors
indicate whether the unbroken symmetry is a pure color, a pure flavor,
or a coupled color-flavor symmetry.

 Our results are summarized in Table 1. The two flavor case is
special. In this case, the dominant order parameter does not
break the color symmetry completely, and the flavor symmetry
is completely unbroken. This is the scenario discussed in
\cite{ARW_98a,RSSV_98}. Subdominant interactions can break
the remaining color symmetry, either with or without \cite{ARW_98a}
flavor symmetry breaking.

  The main result is that, for three flavors, we verify that 
color-flavor locking is indeed the preferred order parameter.
We find that all quark species acquire a mass gap, and both
color and flavor symmetry are completely broken. There are
eight combinations of color and flavor symmetries that generate
unbroken global symmetries. These are the generators of the
diagonal $SU(3)^{c+L+R}$. Also, the quark mass gaps fall into
representations (8+1) of the unbroken symmetry. And, as mentioned
above, the singlet state is twice heavier than the octet. 

  Note that in the present analysis, which only takes into account 
the leading interaction, the order parameter is completely anti-symmetric
in both color and flavor. We find $\Delta^{\alpha\beta}_{ab} \sim 
\epsilon^{\alpha\beta I}\epsilon_{abI}$. If subleading interactions
are taken into account, the order parameter will have the more
general form $\Delta^{\alpha\beta}_{ab}=\kappa_1\delta^\alpha_a
\delta^\beta_b +\kappa_2\delta^\alpha_b\delta^\beta_a$. This order 
parameter leaves the same residual symmetry.

  The main qualitative results we found for three flavors 
extend to $N_f>3$. Color symmetry is always completely broken, and 
all quarks acquire a mass gap. The only remaining symmetries are global
coupled color-flavor symmetries. For massless quarks,
chiral symmetry is spontaneously broken. For an odd number of flavors,
there are subleading instanton operators that, after the
dominant gap is formed, can give an expectation value to
operators of the form $\bar\psi_L\psi_R$. For even numbers
of flavors, generating a non zero $\langle\bar\psi_L\psi_R
\rangle$ is more subtle. Instantons can only give an
expectation value to operators of the form $(\bar\psi_L\psi_R)^2$.

  $N_f=N_c$ (or a multiple thereof) is the most favorable case, in 
the precise sense that in this case the condensation energy
per species is maximal. If $N_f$ is a multiple of $N_c$, the
dominant gap corresponds to multiple embeddings of the
$N_f=N_c$ order parameter. We have not studied the case $N_c\neq
3$ systematically, but since color and flavor are interchangeable 
in (\ref{laa}), the case $N_f =3$ for various numbers of flavors 
is covered implicitly. Also, we have
not studied the interesting case $N_f=N_c\to\infty$.

\begin{table}
\begin{tabular}{|c|c|c|l|c|c|c|}\hline
$N_c$ & $N_f$ & $N_{par}$  &  gaps (deg)  &  $\Delta$
    & $-\epsilon/(N_cN_f)$  &   $N_{sym}$  \\ \hline\hline
  3   &   2   &     3      & $\Delta$ (4)
&   $\Delta_0$   & $\epsilon_0$ &  3 (fl) + 3 (col) \\  \hline
  3   &   3   &     9      & $\Delta$ (8), $2\Delta$ (1)
&   $0.80\Delta_0$ & $1.27\epsilon_0$   &  8  \\   \hline
  3   &   4   &    18      & $\Delta$ (8), $2\Delta$ (4)
&   $0.63\Delta_0$ & $1.21\epsilon_0$  &  6  \\    \hline
  3   &   5   &    30      & $\Delta$ (5), $2\Delta$ (7), $3\Delta$ (3)
&   $0.43\Delta_0$ & $1.18\epsilon_0$  &  3  \\    \hline
  3   &   6   &    45      & $\Delta$ (16), $2\Delta$ (2)
&   $0.80\Delta_0$ & $1.27\epsilon_0$  &  9  \\ \hline
\end{tabular}
\caption{Groundstate properties of the s-wave superfluid
state in QCD with $N_c=3$ colors and $N_f$ flavors. $N_{par}
=N_c(N_c-1)N_f(N_f-1)/4$ is the number of totally anti-symmetric
gap parameters. The column labeled ``gaps (deg)'' gives the
relative magnitude of the gaps in the fermion spectrum, together
with their degeneracy. The numerical values of the gap and
the condensation energy per species are given in units of
$\Delta_0=36\,{\rm MeV}$ and $\epsilon_0=0.73\,{\rm MeV}/{\rm
fm}^3$, respectively. $N_{sym}$ is the number of unbroken
color-flavor symmetries. }

\end{table}

Acknowledgement: We wish to thank S. Treiman for some helpful questions.

\end{document}